\documentstyle[aps,prb,multicol]{revtex}

\begin{document}
\draft

\title{Effects of Al doping on the structural and electronic properties
       of Mg$_{1-x}$Al$_x$B$_2$}

\author{O. de la Pe\~na, A. Aguayo, and R. de Coss\cite{e-mail}}

\address{Departamento de F{\'\i}sica Aplicada, Centro de Investigaci\'on y de Estudios Avanzados\\
        Apartado Postal 73 Cordemex 97310 M\'erida, Yucat\'an, M\'exico}

\date{\today}
\maketitle

\begin{abstract} 
We have studied the structural and electronic properties of
${\rm Mg}_{1-x}{\rm Al}_x{\rm B}_2$ within the Virtual Crystal Approximation (VCA) 
by means of first-principles total-energy calculations. Results for the lattice 
parameters, the electronic band structure, and the Fermi surface as a function of 
Al doping for $0\le x\le 0.6$ are presented. The {\it ab initio} VCA calculations
are in excellent agreement with the experimentally observed change in the lattice 
parameters of Al doped ${\rm MgB}_2$. The calculations show that the Fermi surface
associated with holes at the boron planes collapses gradually with aluminum doping 
and vanishes for $x=0.56$. In addition, an abrupt topological change in the 
$\sigma$-band Fermi surface was found for $x=0.3$. The calculated hole density 
correlates closely with existing experimental data for $T_c(x)$, indicating that 
the observed loss of superconductivity in ${\rm Mg}_{1-x}{\rm Al}_x{\rm B}_2$ is a 
result of hole bands filling.
\end{abstract}

\pacs{PACS: 74.25.Jb, 74.62.Dh, 74.70.Ad, 74.70.Dd}

\begin{multicols}{2}


The discovery of superconductivity in the simple binary intermetallic 
compound ${\rm MgB}_2$ with a $T_c$ as high as 40 K\cite{akimitsu} has 
stimulated intense investigations, both from the experimental and the 
theoretical points of view\cite{review}. The superconducting transition 
temperature $T_c$ for ${\rm MgB}_2$ has been studied as a function of 
pressure and alloying\cite{review}. Pressure studies have shown that $T_c$ 
decreases with applied hydrostatic pressure\cite{monte,prassides,bordet}, 
which has been explained by an increase of the band filling of the boron 
$\sigma$-bands with pressure\cite{prassides,bordet}. Thus, the change in 
the band filling in ${\rm MgB}_2$ under pressure is an effect of the 
reduction of the cell volume with pressure\cite{prassides,bordet}.

Experimentally it has been observed that the superconducting transition 
temperature of ${\rm Mg}_{1-x}{\rm Al}_x{\rm B}_2$ decrease with Al 
doping \cite{cava,bianconi}, and superconductivity disappears for 
$x>0.5$ \cite{postorino,renker}. 
According to band structure calculations of ${\rm MgB}_2$, electron doping 
reduces the density of states (DOS) at the Fermi level\cite{pickett,kortus}. 
Based on the Rigid Band Approximation, An and Pickett\cite{pickett} analyzed 
the effect of Al doping on the DOS of ${\rm MgB}_2$, and found that the DOS 
at the Fermi level drops for $x\approx 0.25$. Structural characterization 
of ${\rm Mg}_{1-x}{\rm Al}_x{\rm B}_2$ shows that the cell volume also
decreases with Al doping\cite{cava,bianconi}. Therefore, the $\sigma$-band 
filling in ${\rm Mg}_{1-x}{\rm Al}_x{\rm B}_2$ is expected to have two 
contributions: the first is due to the electron doping, and the second is 
a result of the cell volume reduction as in the case of pressure effects.

Measurements of the thermoelectric power, $S$, on ${\rm Mg}_{1-x}{\rm Al}_x
{\rm B}_2$ for $x\le 0.1$ have shown that the slope of the linear part 
of $S(T)$ changes with Al doping, indicating changes in the Fermi 
surface due to electron doping\cite{lorenz}. More recently, it was shown
that the Raman spectra of ${\rm Mg}_{1-x}{\rm Al}_x{\rm B}_2$ for
$0\le x\le 0.6$\cite{postorino,renker} show a pronounced frequency shift 
and a considerable change in the line-width for the $E_{2g}$ phonon mode 
at $x\approx 0.3$, which correlate with a steeping in the behavior of 
$T_c(x)$ with Al doping\cite{cava,postorino,renker}. Although some of the 
observed effects of Al doping in ${\rm Mg}_{1-x}{\rm Al}_x{\rm B}_2$ can be 
interpretated qualitatively in terms of the Rigid Band Approximation as 
an effect of the $\sigma$-band filling, a quantitative analysis is
essential in order to determine the interplay between electron doping and 
the structural, electronic, and transport properties of ${\rm Mg}_{1-x}
{\rm Al}_x{\rm B}_2$. 

In this paper, we present a study of the effects of Al doping on the 
structural and electronic properties of ${\rm Mg}_{1-x}{\rm Al}_x{\rm B}_2$ 
for $0\le x\le 0.6$, using the {\it ab initio} Virtual Crystal Approximation. 
The calculated lattice parameters are compared with available experimental 
data\cite{cava}. The evolution of the electronic band structure and the 
$\sigma$-band Fermi surface (FS) as a function of Al doping is analyzed. 
We correlate our results to the experimentally observed behavior of 
$T_c$ with Al doping\cite{cava,bianconi,postorino,renker}.
We show that the observed loss of superconductivity in ${\rm Mg}_{1-x}
{\rm Al}_x{\rm B}_2$ can be explained by the filling of the hole bands.


The Khon-Sham total energies were calculated self-consistently using the 
full-potential linearized augmented plane-wave method (LAPW)\cite{lapw}
as implemented in the WIEN97 code\cite{wien}, where the core states are 
treated fully relativistically, and the semicore and valence states are 
computed in a scalar relativistic approximation. The exchange correlation 
potential was evaluated within the generalized gradient approximation 
(GGA), using the recent parameter-free GGA form by Perdew, Burke, and 
Ernzerhof\cite{pbe}. We chose muffin-tin radii $(R_{MT})$ of 1.8 and 
1.5 a.u. for Mg and B, respectively, and used a plane-wave cutoff 
$R_{MT}K_{MAX}=9.0$. Inside the atomic spheres the potential and charge 
density were expanded in crystal harmonics up to $l=10$. Convergence 
was assumed when the energy difference between the input and output 
charge densities was less than $1\times 10^{-5}$ Ryd. Special attention 
was paid to convergence of results by performing the calculations for a 
sufficiently large number of $k$ points in the irreducible wedge of the 
Brillouin zone for the omega structure (144 $k$ points). The corrected
tetrahedron method was used for Brillouin-zone integration\cite{blochl}.

The Al doping was modeled in the {\it ab initio} Virtual Crystal 
Approximation (VCA)\cite{vca1,vca2}. The Mg $(Z=12)$ sites are substituted 
by pseudo-atoms which have a fractional electronic charge $(Z=12+x)$, 
depending on the Al concentration, $x$. This approximation is justified 
mainly by the fact that Al only has one electron more than Mg. The 
full-potential for the VCA system it is determined self-consistently 
for each value of Al doping without shape approximation\cite{wien}. 
The {\it ab initio} VCA as implemented in this work has been used very 
recently to model C, Cu and Be substitutions in MgB$_2$\cite{vca2}. 
The equilibrium lattice parameters were determined by total-energy 
calculations for each value of Al doping ($x$=0.0, 0.1, 0.2, 0.25, 0.3, 
0.35, 0.4, 0.5, and 0.6). Since the AlB$_2$ (omega) structure has 
two structural parameters ($a$ and $c$), we performed self-consistent 
total-energy calculations for nine different volumes and for nine
different $c/a$ ratios, in order to optimize both $V$ and $c/a$ for 
each Al concentration. For MgB$_2$ we have obtained $a=3.083$ {\AA} and 
$c=3.526$ {\AA}, which compares very well to the experimental values of 
$a=3.086$ {\AA} and $c=3.524$ {\AA} \cite{akimitsu}, respectively.


In Fig. 1 we present the calculated lattice parameters ($a$ and $c$) 
of ${\rm Mg}_{1-x}{\rm Al}_ {x}{\rm B}_2$ for $0\le x\le 0.6$, and it
can be seen that both $a$ and $c$ decrease with Al doping as observed 
experimentally\cite{cava,bianconi}. For comparison we have included the 
experimental data from Slusky {\it et al.}\cite{cava}, and we find that the 
change in $a$ with Al doping is very well reproduced by the VCA calculations. 
For the region $0.1\le x\le 0.2$ two values of $c$ were reported\cite{cava},
which has been ascribed to the coexistence of two phases (Mg-rich and 
Al-rich phases). It is interesting to note that in this region the VCA values 
approximately reproduce the average value. However, for $x>0.2$ we find very 
good agreement between the experimental data and the VCA calculations (see 
Fig. 1). Although both cell parameters ($a$ and $c$) decrease monotonically 
with Al doping, it is interesting to note that the slope for $c$ as a 
function of Al doping is larger than for $a$. 
In order to understand this behavior in the context of bonding properties
we have analyzed the change in the charge distribution with Al doping. 
Fig. 2(a) shows the charge density distribution in the (110) plane of 
MgB$_2$. Mg nuclei are located at the corners of the map and B nuclei are 
at the $(1/3,1/2)$ and $(2/3,1/2)$ positions, all of them in the plane of
the figure. We can see the directional, covalent B-B $\sigma$-bonds.
In addition, there is a significant density of charge in the interstitial
region giving rise to metallic-type bonding between the Mg and B planes.
The charge distribution and bonding properties of MgB$_2$ have been 
calculated previously and were discussed in detail in Ref. 19.
Therefore, we concentrate on the influence of Al doping on the electron 
density of ${\rm MgB}_2$. Fig. 2(b) shows the difference between the charge 
densities of ${\rm Mg}_{0.5}{\rm Al}_{0.5}{\rm B}_2$ and ${\rm MgB}_2$. It can 
be seen that charge transfer occurs from Al-B ions into the nearby interstitial
region. We can see that the majority of this charge is distributed in the 
inter-plane region, and an important fraction of the charge is being transfered 
to the $\pi$-bond, while only a small fraction is at the $\sigma$-bond in the boron 
planes. This important increase in occupation of the $\pi$-bond with Al
doping accounts for the strong decrease of the separation between planes 
(the $c$ axis) of ${\rm Mg}_{1-x}{\rm Al}_x{\rm B}_2$.

The evolution of the calculated DOS for 
${\rm Mg}_{1-x}{\rm Al}_x{\rm B}_2$, not shown here, shows a band 
broadening as a function of Al doping, mainly as a consequence of the 
cell volume reduction. In addition, electron doping raises the Fermi 
level to higher energies. Both effects, band broadening and electron 
doping, contribute to reduce the density of states at the Fermi level 
in Al-doped ${\rm MgB}_2$. In Table I, we summarize the calculated lattice 
parameters, cell volume, and the total density of states at the Fermi 
level [$N(E_F)$] for each of the studied Al concentrations. We can see that 
$N(E_F)$ decreases with Al doping, from 0.72 for ${\rm MgB}_2$ to 0.26 
states/eV per cell for ${\rm Mg}_{0.4}{\rm Al}_{0.6}{\rm B}_2$. 
Therefore, in a BCS scenario this reduction in $N(E_F)$ accounts for 
the decrease of $T_c$ with Al doping in 
${\rm Mg}_{1-x}{\rm Al}_x{\rm B}_2$\cite{cava,bianconi,postorino,renker}.

A careful analysis of the $x$-dependence of the electronic band 
structure, and in particular of the $\sigma$-band FS which has been
shown to be relevant for superconductivity in 
${\rm MgB}_2$\cite{pickett,kortus,kong,bohnen,yildirim,amy}, provides a 
more detailed and quantitative description of the effects of Al doping. 
In Fig. 3 we present the electronic band structure of ${\rm MgB}_2$. 
The $\sigma$-bands coming from the $s$-$p$ boron orbitals, are 
strongly two-dimensional with very little dispersion along $\Gamma$-$A$, 
this dispersion can be characterized by the difference between the 
$E_{\Gamma}$ and $E_A$ energies (see Fig. 3). The $E_{\Gamma}$ and $E_A$ 
energies correspond to the bottom and top of the $\sigma$-band in the 
$\Gamma$-$A$ direction, respectively. The light-hole and heavy-hole 
$\sigma$-bands in ${\rm MgB}_2$ form a FS consisting of two fluted 
cylinders surrounding the $\Gamma$-$A$ line in the Brillouin zone (see Fig. 3). 
The dependence of the energy of the $\sigma$-bands at $\Gamma$ and $A$ 
relative to $E_F$ as a function of Al doping are shown in Fig. 4. We can 
see that both energies, $E_{\Gamma}$ and $E_A$, decrease monotonically 
as a function of Al doping. 
More interestingly, the Fermi level reaches $E_{\Gamma}$ for $x=0.3$ 
and $E_A$ for $x=0.56$. We find that the radius of the cylinders 
decreases gradually with Al doping and at a critical concentration 
of $x=0.3$, the radius at $k_z=0$ collapses and the FS takes 
the form of a sandglass. A three-dimensional view of the changes in 
the FS topology with Al doping are presented in Fig. 4. For $x=0.3$ 
the Fermi level in $\Gamma$ is at a saddle point in the band structure, 
and the transition through the saddle point results in the disruption 
of the neck, i.e., the transition from a closed to an open section 
of FS\cite{chu}. For $x>0.3$ the FS takes the form of two cones 
(see Fig. 4), and these finally vanishes at the second critical 
concentration $(x=0.56)$ when the hole bands have been filled. These 
changes in the hole FS are expected to be accompanied by various kinds 
of electronically driven anomalies, including lattice dynamics and 
transport properties.

As was mentioned above, Raman spectroscopy studies on ${\rm Mg}_{1-x}
{\rm Al}_x{\rm B}_2$ \cite{postorino,renker} show a pronounced shift 
and a considerable change in the line-width of the $E_{2g}$ phonon 
mode at $x\approx 0.3$. Additionally, a steeping of the $T_c$ 
decrease has been observed at an Al concentration of approximately 
0.3\cite{cava,postorino,renker}. These changes in both the structural 
and the superconducting properties seems to be strongly related to the 
abrupt change in the FS topology, which occurs for $x=0.3$ (see Fig. 4). 
In order to establish a more direct comparison between the FS evolution 
and the superconducting properties with increasing the Al doping,
we have calculated the hole FS area as a function of $x$, which is 
proportional to the hole density at the Fermi level. In Fig. 4(b)
we compare the calculated normalized FS area, $A_{FS}(x)/A_{FS}(0)$, 
with the normalized superconducting critical temperaure, $T_c(x)/T_c(0)$.
The experimental data for $T_c(x)$ were taken from Ref. 8. 
We can see that for the low concentration region ($x\le 0.25$), before
the $E_{2g}$ phonon frequency shift\cite{renker}, the drop of $T_c$ is 
directly related to the change in the hole density. This view is in
agreement with recent results of NMR experiments on Al-doped ${\rm MgB}_2$ 
for $x\le 0.1$\cite{kote}. In the high concentration region $(x>0.25)$, 
the behavior of $T_c$ is determined by the FS area but the importance 
of the phonon-renormalization is clear\cite{postorino,renker}. 
In this way, the FS area and $T_c$ follow the same behavior with Al 
doping in the whole range $(0\le x\le 0.6)$, indicating a close 
relation between the changes in the $\sigma$-band FS and the loss of 
superconductivity in Al doped ${\rm MgB}_2$. 


In summary, we have performed a first-principles study of the effects 
of Al doping on the structural parameters, the electronic structure, 
and the $\sigma$-band FS of ${\rm Mg}_{1-x}{\rm Al}_{x}{\rm B}_2$, 
using the Virtual Crystal Approximation. 
(i) We find that the {\it ab initio} VCA calculations are in excellent
agreement with the experimentally observed changes in the lattice 
parameters as a function of Al doping. 
(ii) The analysis of the charge density shows that an important portion of 
the Al-electrons are at the inter-plane region and only a small fraction at 
the B-B planes, providing an explanation for the strong change of the 
$c$-axis and the small change in the $a$-axis with the Al concentration.
(iii) The hole FS gradually collapses with Al doping and vanishes for 
$x=0.56$. An abrupt topological change was found for $x=0.3$, which 
correlates with the frequency shift of the $E_{2g}$ phonon mode and the 
steeping in the $T_c(x)$ decrease. Additionally, the critical concentration 
of $x=0.56$ at which the hole FS disappears, corresponds to the experimentally 
observed Al concentration (0.5-0.6) for which $T_c(x)$ vanishes.
(iv) We find that the behavior of the calculated $\sigma$-band FS area with 
Al doping correlates with the superconducting critical temperature $T_c(x)$.
Consequently, the observed loss of superconductivity in ${\rm Mg}_{1-x}
{\rm Al}_x{\rm B}_2$, can be explained as a result of the filling of the 
hole bands.

This research was funded by the Consejo Nacional de Ciencia y Tecnolog{\'\i}a 
(CONACYT, M{\'e}xico) under Grant No. 34501-E. Two of the authors (O.P. and 
A.A.) gratefully acknowledge a student fellowship from CONACYT-M\'exico. 
The authors would like to thank Dimitris Papaconstantopoulos and David Singh 
for valuable discussions. 



\begin{figure}
\label{fig1} 
\caption{Lattice parameters $a$ and $c$ for ${\rm Mg}_{1-x}{\rm Al}_x
{\rm B}_2$. Experimental data from Ref. [6] (solid circle) and calculated 
using VCA (open circle).}
\end{figure}

\begin{figure}
\label{fig2} 
\caption{Electronic charge density for the ${\rm MgB}_2$(110) plane (top) 
and the charge density difference between ${\rm Mg}_{0.5}{\rm Al}_{0.5}
{\rm B}_2$ and ${\rm MgB}_2$ (bottom).}
\end{figure}

\begin{figure}
\label{fig3} 
\caption{Electronic band structure and hole Fermi surface for ${\rm MgB}_2$ 
at the calculated lattice constants (see Table I).}
\end{figure}

\begin{figure}
\label{fig4} 
\caption{(a)Energy position of the $\sigma$-band at $\Gamma$ ($E_{\Gamma}$)
and $A$ ($E_A$) relative to $E_F$ for ${\rm Mg}_{1-x}{\rm Al}_x{\rm B}_2$. 
In the inset, the hole Fermi surface for $x=0.25$, 0.3, and 0.35. 
(b) Calculated normalized holes FS area, $A(x)/A(0)$ (solid line) and the
normalized experimental values (Ref.[8]) of the superconducting 
critial temperature, $T_c(x)/T_c(0)$ (open circles).}
\end{figure}

\begin{table}
\label{tab1}
\caption{Calculated lattice parameters, cell volume, and density of 
states at the Fermi level, $N(E_F)$ in states/eV unit cell, for 
${\rm Mg}_{1-x}{\rm Al}_x{\rm B}_2$ as a function of Al doping $(x)$.}
\begin{tabular}{lcccc}
$x$ & $a$(\AA) & $c$(\AA) & $V$({\AA}$^3$) & $N(E_F)$ \\
\tableline
0.0  & 3.083 & 3.526 & 29.02 & 0.72 \\  
0.1  & 3.076 & 3.486 & 28.56 & 0.68 \\
0.2  & 3.072 & 3.448 & 28.18 & 0.64 \\
0.25 & 3.070 & 3.424 & 27.95 & 0.60 \\
0.3  & 3.063 & 3.403 & 27.65 & 0.55 \\
0.35 & 3.059 & 3.386 & 27.44 & 0.48 \\
0.4  & 3.055 & 3.367 & 27.21 & 0.43 \\
0.5  & 3.047 & 3.338 & 26.84 & 0.33 \\
0.6  & 3.039 & 3.315 & 26.51 & 0.26 
\end{tabular}
\end{table}

\end{multicols}


\begin{references}

\bibitem[*] {e-mail} Author to whom correspondence should be
addressed. Electronic address: decoss@mda.cinvestav.mx

\bibitem{akimitsu} J. Nagamatsu, N. Nakagawa, T. Muranaka, Y. Zenitai, 
and J. Akimitsu, Nature {\bf 410}, 63 (2001).

\bibitem{review} C. Buzea and T. Yamashita, 
Superc. Sci. Technol. {\bf 14}, R115 (2001).

\bibitem{monte} M. Monteverde, M. N{\'u}{\~n}ez-Regueiro, N. Rogado,
K.A. Regan, M.A. Hayward, T. He, S.M. Loureiro, and R.J. Cava,
Science {\bf 292}, 75 (2001).

\bibitem{prassides} K. Prassides, Y. Iwasa, T. Ito, D.H. Chi, K. Uehara,
E. Nishibori, M. Takata, M. Sakata, Y. Ohishi, O. Shimomura, 
T. Muranaka, and J. Akimitsu, Phys. Rev. B {\bf 64}, 012509 (2001).

\bibitem{bordet} P. Bordet, M. Mezouar, M. N{\'u}{\~n}ez-Regueiro, 
M. Monteverde, M.D. N{\'u}{\~n}ez-Regueiro, N. Rogado, K.A. Regan, 
M.A. Hayward, T. He, S.M. Loureiro, and R.J. Cava,
Phys. Rev. B {\bf 64}, 172502 (2001).

\bibitem{cava} J.S. Slusky, N. Rogado, K.A. Regan, M.A. Hayward,
P. Khalifah, T.He, K. Inumaru, S.M. Loureiro, M.K. Hass,
H.W. Zandbergen, and R.J. Cava, Nature {\bf 410}, 343 (2001).

\bibitem{bianconi} A. Bianconi, D. Di Castro, S. Agrestini, G. Campi,
N.L. Saini, A. Saccone, S. De Negri, and M. Giovannini, 
J. Phys.: Condens. Matter {\bf 13}, 7383 (2001).

\bibitem{postorino} P. Postorino, A. Congeduti, P. Dore, A. Nucara,
A. Bianconi, D. Di Castro, S. De Negri, and A. Saccone,
Phys Rev. B {\bf 65}, 020507 (2002).

\bibitem{renker} B. Renker, K.B. Bohnen, R. Heid, D. Ernst, 
H. Schober, M. Koza, P. Adelmann, P. Schweiss, and T. Wolf, 
Phys. Rev. Lett. {\bf 88}, 067001 (2002).

\bibitem{pickett} J.M. An and W.E. Pickett, 
Phys. Rev. Lett. {\bf 86}, 4366 (2001).

\bibitem{kortus} J. Kortus, I.I. Mazin, K.D. Belashchenko,
V.P. Antropov, and L.L. Boyer, Phys. Rev. Lett. {\bf 86}, 4656 (2001).

\bibitem{lorenz} B. Lorenz, R.L. Meng, Y.Y.Xue, and C.W. Chu,
Phys. Rev. B {\bf 64}, 052513 (2001).

\bibitem{lapw} D.J. Singh, {\it Plane Waves, Pseudopotentials and the
LAPW Method} (Kluwer Academic Publishers, Boston, 1994).

\bibitem{wien} P. Blaha, K. Schwarz, and J. Luitz, computer code 
WIEN97(Vienna University of Technology, 1997), improved and updated 
Unix version of the original copyrighted WIEN code, wich was
published by P. Blaha, K. Schwarz, P. Sorantin, and S.B. Trickey,
Comput. Phys. Commun. {\bf 59}, 339 (1990).

\bibitem{pbe} J.P. Perdew, S. Burke, and M. Ernzerhof,
 Phys. Rev. Lett. {\bf 77}, 3865 (1996). 

\bibitem{blochl} P.E. Bl{\"o}chl, O. Jepsen, and O.K. Andersen,
Phys. Rev. B {\bf 49}, 16223 (1994).

\bibitem{vca1} D.A. Papaconstantopoulos, E.N. Economou, B.M. Klein,
and L.L. Boyer, Phys. Rev. B {\bf 20}, 177 (1979).

\bibitem{vca2} M.J. Mehl, D.A. Papaconstantopoulos, and D.J. Singh,
Phys. Rev. B {\bf 64}, (2001).

\bibitem{bond}K.D. Belashchenko, M. van Schilfgaarde, V.P. Antropov,
Phys. Rev. B {\bf 64}, 092503 (2001); P. Ravindran, P. Vajeeston, R. Vidya,
A. Kjekshus, and H. Fjellvag, Phys. Rev. B {\bf 64}, 224509 (2001).

\bibitem{kong} Y. Kong, O.V. Dolgov, O. Jepsen, and O.K. Andersen,
Phys. Rev. B {\bf 64}, 020501 (2001).

\bibitem{bohnen} K.P. Bohnen, R. Heid, and B. Renker,
Phys. Rev. Lett. {\bf 86}, 5771 (2001).

\bibitem{yildirim} T. Yildirim, O. G\"ulseren, J.W. Lynn, C.M. Brown,
T.J. Udovic, Q. Huang, N. Rogado, K.A. Regan, M.A. Hayward, J.S. Slusky,
T. He, M.K. Hass, P. Khalifah, K. Inumaru, and R.J. Cava,
Phys. Rev. Lett. {\bf 87}, 37001 (2001).

\bibitem{amy} A.Y. Liu, I.I. Mazin, and J. Kortus,
Phys. Rev. Lett. {\bf 87}, 87005 (2001).

\bibitem{chu} C.W. Chu, T.F. Smith, and W.E. Gardner,
Phys. Rev. B {\bf 1}, 214 (1970).

\bibitem{kote} H. Kotegawa, K. Ishida, Y. Kitaoka, T. Muranaka, 
N. Nakagawa, H. Takagiwa, and J. Akimitsu, cond-mat/0201578.

\end{references}
\end{document}